\documentclass[technote,onecolumn,10pt]{IEEEtran}
\usepackage[english]{babel}
\usepackage{graphicx}
\usepackage[margin=2cm]{geometry}
\usepackage{float}

\newcommand{\mathd}{\mathrm{d}}
\newcommand{\mathe}{\mathrm{e}}
\newcommand{\nospace}{}
\newcommand{\tmtextbf}[1]{{\bfseries{#1}}}
\newcommand{\tmtextit}[1]{{\itshape{#1}}}

\begin{document}

\title{Spaghetti prediction: A robust method for forecasting short time
series}

\author{
  \IEEEauthorblockN{Steven C. Gustafson$^1$ and Leno M. Pedrotti$^2$}\\
  \IEEEauthorblockA{$^1$\tmtextit{Retired faculty, Air Force Institute of
  Technology, Dayton, Ohio, gustafson.steven @ gmail.com}\\
  $^2$\tmtextit{Department of Physics, University of Dayton, Dayton, Ohio,
  lpedrotti1 @ udayton.edu}}
}

\maketitle

\begin{abstract}
  A novel method for predicting time series is described and demonstrated.
  This method inputs time series data points and outputs multiple
  ``spaghetti'' functions from which predictions can be made. Spaghetti
  prediction has desirable properties that are not realized by classic
  autoregression, moving average, spline, Gaussian process, and other methods.
  It is particularly appropriate for short time series because it allows
  asymmetric prediction distributions and produces prediction functions which
  are robust in that they use multiple independent models.
\end{abstract}

\section{Motivation}

Televised hurricane forecasts often feature plots that appear as multiple
strands of cooked spaghetti tossed on a plate. Such ``spaghetti'' plots
indicate the hurricane path predicted by independent meteorological models.
Similarly, the novel spaghetti time series prediction method described and
demonstrated here develops independent spaghetti functions, each of which
constitutes a model of given time series data.

\section{Description}

Given time series points $(x_i, y_i)$ with $i = 1, 2, \ldots n$, the spaghetti
functions are
\begin{equation}
  f_i (x) = a_i + b_i x + \sum_k A_{i \nospace k} \mathe^{(x - x_k)^2 / (2
  \sigma_i^2)},
\end{equation}
where $k$ is an index over the points with point $i$ left out, $a_i + b_i x$
is the least-squares line of the $(x_k, y_k)$, $A_{i \nospace k}$ and
$\sigma_i$ are chosen to minimize
\[ \sum_k [y_k - f_i (x_k)]^2 + \lambda_i  \int_{- \infty}^{\infty} \left(
   \frac{\mathd^2 f_i (x)}{\mathd x^2} \right)^2 \mathd x, \]
and $\lambda_i$ is positive and minimizes $| y_i - f_i (x_i) |$.\\

Comments on this description are as follows:
\begin{enumerate}
  \item Each spaghetti function is developed with one time series point left
  out, so there are $n$ functions. Each function is the least squares line of
  the $n - 1$ points plus a weighted linear combination of Gaussian (i.e.,
  normal distribution) basis functions or kernels, where each kernel is
  centered on one of the points and has the same variance.
  
  \item Each spaghetti function has its kernel weights and kernel variance
  determined so that deviation plus the product of a positive constant and
  roughness is minimized. Deviation has the classic sum squared difference
  from the points definition, and roughness has the classic integrated squared
  second derivative definition [MacKay, 2003; Gustafson et al., 2007].
  
  \item If $\lambda_i$ is nearly zero, then deviation minimization dominates,
  and the spaghetti function nearly intersects the $n - 1$ points. If
  $\lambda_i$ is nearly infinite, then roughness mnimization dominates, and
  the spaghetti function is nearly the least squares line of the $n - 1$
  points. Thus $\lambda_i$ controls the function shape in a manner consistent
  with classic (i.e., standard or well-known) regularization methods [MacKay,
  2003; Gustafson et al., 2011].

  \item The $\lambda_i$ for each spaghetti function is determined so as to
  best predict the left out point, i.e., it is such that the absolute
  difference between the function and the left out point is minimized. Thus
  the shape of each spaghetti function is adjusted so that it predicts the
  left out point with least error. \tmtextit{This method for developing
  multiple independent models is novel in that (to our knowledge) it has not
  been reported in the technical literature.}
\end{enumerate}
Two classic functions may be compared with the spaghetti functions:
\begin{enumerate}
  \item The least squares line of all of the points: $g (x) = a + bx$, where
  $a$ and $b$ minimize $\sum_i [y_i - g (x_i)]^2$. Note that $g (x)$ is the
  spaghetti function of all the points as $\lambda$ approaches infinity and
  that it is also the function of zero roughness that has least deviation.
  
  \item The least rough interpolator of all the points: $h (x) = a + bx +
  \sum_i A_i \exp [(x - x_i)^2 / (2 \sigma^2)]$, where the $A_i$ are such that
  $h (x_i) = y_i$ and $\sigma$ is such that $\int_{- \infty}^{\infty} \left(
  \frac{\mathd^2 h (x)}{\mathd x^2} \right)^2 \mathd x$ is minimized. Note
  that $h (x)$ is the spaghetti function of all the points as $\lambda$
  approaches zero and that it is also the function of zero deviation that has
  least roughness.
\end{enumerate}

\section{Demonstration}

Figure 1 shows seven time series points, the seven spaghetti functions $f_i
(x)$, and the mean $\mu (x)$ of the spaghetti functions. As described above,
each spaghetti function has its deviation-roughness tradeoff optimized so that
it predicts its corresponding left out point with least error. The spaghetti
functions constitute multiple independent models of the time series data and
enable robust prediction at any time. In general, the distribution of
spaghetti function predictions at any time is not symmetric. Fiigure 1
indicates this asymmetry in that that the mean of the spaghetti functions
differs from the median function (i.e., the middle function). Asymmetric
prediction distributions are particularly realistic for short time series.

\begin{figure}[H]
\centering
  \includegraphics[scale=0.9]{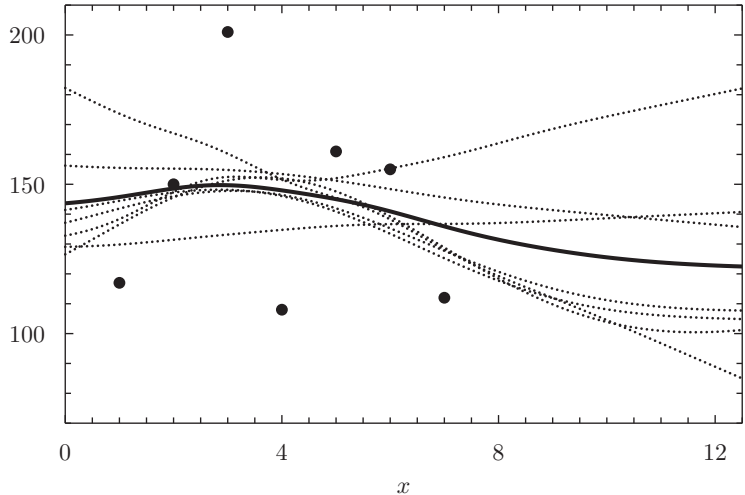}
  \caption{Seven time series points (dots), the seven spaghetti functions $f_i
  (x)$ (dotted curves), and the mean $\mu (x)$ (solid curve) of the spaghetti
  functions. Note that the distribution of spaghetti functions is asymmetric
  in that the median or middle spaghetti function differs from $\mu (x)$.}
\end{figure}

Figure 2 shows the seven points, $\mu (x)$, $\mu (x) + s (x)$, and $\mu (x) -
s (x)$, where $s (x)$ is the standard deviation of the spaghetti functions.~
The standard deviation of the spaghetti functions should be small near the
points and large away from them, because prediction should be more accurate
near the points. The figure shows that this behavior is realized.

\begin{figure}[H]
  \centering
  \includegraphics[scale=0.9]{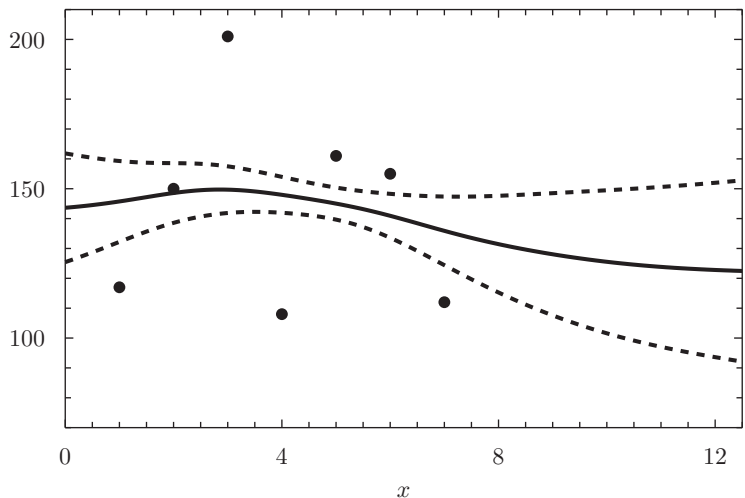}
  \caption{The seven points (dots), $\mu (x)$ (solid curve), $\mu (x) + s (x)$
  (dashed curve), and $\mu (x) - s (x)$ (dashed curve), where $s (x)$ is the
  standard deviation of the spaghetti functions. Note that $s (x)$ is small
  near the points and large away from them.}
\end{figure}

Figure 3 shows the seven points and three functions that characterize them:
their least squares line $g (x)$, their least rough interpolator $h (x)$, and
the mean of the spaghetti functions $\mu (x)$. The mean of the spaghetti
functions should extrapolate to a line, because prediction should employ the
simplest functional form far from the points. The figure illustrates this
behavior, and it shows that the line is the least squares line of the points.

\begin{figure}[H]
  \centering
  \includegraphics[scale=0.9]{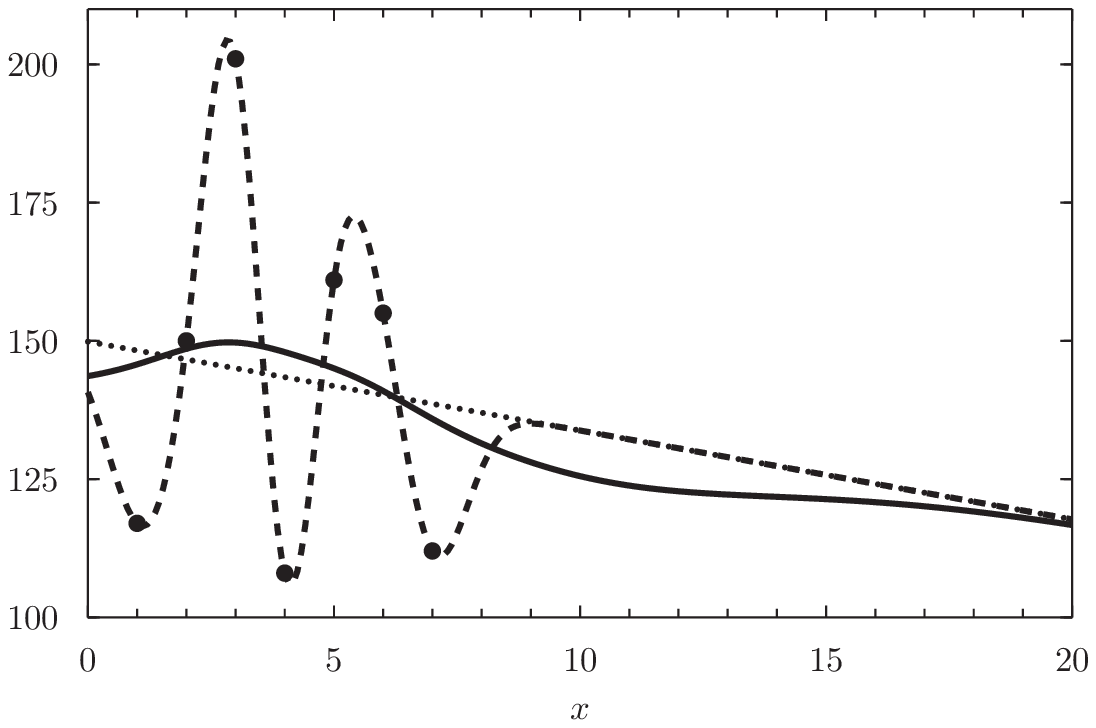}
  \caption{The seven points and three functions that characterize them:~ their
  least squares line $g (x)$ (dotted curve), their least rough interpolator $h
  (x)$ (dashed curve), and the mean of the spaghetti functions $\mu (x)$
  (solid curve). Note that $\mu (x)$ extrapolates to $g (x)$.}
\end{figure}

\section{Discussion}

Spaghetti prediction could be compared with numerous time series prediction
methods on various real or synthetic data sets. However, as is well known,
data can almost always be found or synthesized (there are only trivial
exceptions) to show that a given prediction method outperforms competing
methods in accord with given criteria [Duda et al., 2000; Poggio et al. 2004].
Thus the focus of comparison should be on the properties of the prediction
methods.

\

A wide range of time series prediction methods [Anava et al., 2013] are in
effect realizations of Gaussian processes that correspond to specific choices
of a covariance function [Rasmussen et al., 2007]. Classic methods, including
autoregressive and moving average methods and their various combinations and
extensions, e.g., the well-known Box-Jenkins methods, are such in-effect
realizations, as are various spline methods, e.g., smoothing and B-splines
[Brockwell, et al. 2009]. However, spaghetti prediction is not a realization
of a Gaussian process. In a Gaussian process the dependent variable has a
Gaussian probability density function (pdf) at any given time, and thus the
pdf is symmetric and the mean, median, and mode are identical. In spaghetti
prediction the pdf of the dependent variable at any given time does not have
this constraint. This property is important for short time series (i.e.,
series that have a small number of given points), which are likely to be
characterized by asymmetric prediction distributions.

\

Spaghetti prediction has other properties that contrast with Gaussian process
methods. A Gaussian process produces a single prediction function and a
variance for this function at any time, but spaghetti prediction produces as
many prediction functions as there are data points. Furthermore, each
spaghetti function has the tradeoff between its roughness and its deviation
from the data with one point left out optimized so as to predict the left out
point with least error. In making this tradeoff, classic definitions of
deviation and roughness are employed, and each spaghetti function has a
classic form: the least squares line of the retained points plus a weighed sum
of point-centered Gaussian kernels of the same variance.~

\

Spaghetti prediction is particularly appropriate for short time series. In
this case a robust method, i.e., a method that produces multiple independent
predictions, is especially important. Also, as discussed above, the
distribution of predictions is typically skewed for small data sets, and thus
a method that permits an asymmetric distribution of predictions is
particularly important. Spaghetti prediction, unlike Gaussian process methods,
has these properties. \

\

\noindent \tmtextbf{References}
\smallskip 

\setlength{\parindent}{0cm}
\small{

O. Anava, E. Hazan, S. Mannor, O. Shamir, ``Online learning for time series
prediction'', arXiv: 1302.6927v1, 2013.

\smallskip 

P. Brockwell, R. Davis, ``Time series: theory and methods'', Springer, 2009.
\smallskip 

R. Duda, P. Hart, D. Stork, ``Pattern classification'', Wiley, 2000.

\smallskip 

S. Gustafson, D. Parker, ``Cardinal interpolation'', IEEE Trans. PAMI 29:
1538 -- 1545, 2007.

\smallskip 

S. Gustafson, A. Hillier, ``A probability density for modeling unknown
physical processes'', arXiv: 1104.3992, 2011.

\smallskip 

D. MacKay, ``Information theory, inference, and learning algorithms'',
Cambridge, 2003.

\smallskip 

T. Poggio, R. Rifkin, S. Mukherjee, P. Niyogi, ``General conditions for
predictivity in learning theory'', Nature 428: 419 -- 422, 2004.

\smallskip 

C. Rasmussen, C. Williams, ``Gaussian processes for machine learning'', MIT
press, 2006.}

\end{document}